\documentclass[prd,showpacs,showkeys,floatfix,nofootinbib,
               preprint,12pt,tightenlines,fleqn,eqsecnum]{revtex4}
\usepackage{amsmath,amssymb,graphicx}

\newcommand  {\version}{v5}

\newcommand{\dd}{\mathrm{d}}                    
\newcommand{\id}{\mathrm{d}}                    
\newcommand{\ii}{\mathrm{i}}                    
\newcommand{\action}{\mathcal{S}}               
\newcommand{\BigO}{\mathrm{O}}                  
\newcommand{\abs}[1]{|{#1}|}                    
\newcommand{\plus}{\oplus}                      
\newcommand{\minus}{\ominus}                    
\newcommand{\diag}{\ensuremath{\mathrm{diag}}}  
\renewcommand{\vec}[1]{\boldsymbol{\mathrm{#1}}}
\newcommand{\sgn}{\mathrm{sgn}}                 

\newcommand{\kmax}{k_\text{max}}
\newcommand{\abskmax}{\kmax}
\newcommand{\absqp}{\abs{q_\parallel}}
\newcommand{\arcsinh}{\mathrm{arcsinh}}

\begin{document}

\noindent Physical Review D 76, 025024 (2007)  \hfill
arXiv:0704.3255 [hep-th] (\version)\newline\vspace*{1\baselineskip}

\title{Vacuum Cherenkov radiation in spacelike Maxwell--Chern--Simons theory}
\author{C.~Kaufhold}\email{kaufhold@particle.uni-karlsruhe.de}
\author{F.R.~Klinkhamer}
\email{frans.klinkhamer@physik.uni-karlsruhe.de}
\thanks{corresponding author}

\affiliation{Institute for Theoretical Physics,\\
University of Karlsruhe (TH),\\
76128 Karlsruhe, Germany\\}

\begin{abstract}
\vspace*{.25\baselineskip}\noindent
A detailed analysis of vacuum Cherenkov
radiation in spacelike Maxwell--Chern--Simons (MCS) theory is presented.
A semiclassical treatment reproduces the leading terms  of the tree-level
result from quantum field theory. Moreover, certain quantum corrections
turn out to be suppressed for large energies of the charged particle,
for example, the quantum corrections to the classical MCS Cherenkov angle.
It is argued that MCS-theory Cherenkov radiation may, in principle, lead to
anisotropy effects for ultra-high-energy cosmic rays (UHECRs).
In addition, a qualitative discussion of vacuum Cherenkov radiation
from a modified-Maxwell term in the action is  given, together with
UHECR bounds on some of its dimensionless ``coupling constants.''
\end{abstract}

\maketitle

\section{Introduction}
\label{sec:introduction}

A charged particle moving with constant velocity in a macroscopic
medium is known to emit Cherenkov radiation if its velocity $v$
exceeds the phase velocity $v_\text{ph}$ of light in the medium
\cite{Cherenkov1934,Vavilov1934,Cherenkov1937,FrankTamm1934,
Ginzburg1940,Cox1944,Frank1984,Jelley1958,Zrelov1970,Afanasiev2004}.
For a nondispersive isotropic medium with refractive index $n>1$,
the phase velocity of light $v_\text{ph}\equiv\omega/|\vec k|=c/n$
is less than $c$, the velocity of light \emph{in vacuo}.
As $c$ corresponds to the maximum attainable velocity of a charged particle
according to the theory of special relativity \cite{Einstein1905},
it is then possible, for a sufficiently fast particle,
to satisfy the Cherenkov condition $v>v_\text{ph}$ and radiate.

In certain Lorentz-violating theories of photons, the photon
four-momentum $p_\mu =\hbar\,k_\mu\equiv\hbar\, (\omega/c,\vec k)$
can, even in the vacuum, be spacelike
($\omega^2/c^2 - |\vec k|^2 <0$), so that $v_\text{ph}<c$.
This last upper bound on the phase velocity allows for
so-called ``vacuum Cherenkov radiation,'' that is, photon emission
by a charged particle moving sufficiently fast in such a vacuum.
The possibility of vacuum Cherenkov radiation
in generic Lorentz-noninvariant theories has been discussed in, e.g.,
Refs.~\cite{Sommerfeld1904,Beall1970,ColemanGlashow1997}.

In this article, we study vacuum Cherenkov radiation in spacelike
Maxwell--Chern--Simons (MCS) theory
\cite{Carroll-etal1990,AdamKlinkhamer2001,AdamKlinkhamer2003},
continuing our previous work \cite{KaufholdKlinkhamer2006}.
Vacuum Cherenkov radiation in this theory has been studied earlier in
Refs.~\cite{LehnertPotting2004PRL93,LehnertPotting2004PRD70}. Here, we are
interested in the comparison with \emph{standard} Cherenkov radiation in
a macroscopic medium characterized by a refractive index.
In addition, we will pay attention to quantum and spin effects,
which will turn out to be relevant at large energies.
The main focus of this article is theoretical, but applications
to cosmic-ray physics will be briefly considered.

The article is organized as follows. In Sec.~\ref{sec:MCS}, we present old
and new results on vacuum Cherenkov radiation in MCS theory, with the
fixed Chern--Simons vector taken to be purely spacelike. (The rather long
expressions for the decay widths and radiation rates are relegated to
Appendices~\ref{sec:appendix-decay-widths} and
\ref{sec:appendix-radiation-rate-coeff}.) Possible physics applications of
MCS-theory Cherenkov radiation include anisotropy effects for
ultra-high-energy cosmic rays (UHECRs),
as will be discussed in Sec.~\ref{sec:MCS-vacuum-Cherenkov-radiation}. In
Sec.~\ref{sec:standard-Cherenkov}, we review certain well-known results on
standard Cherenkov radiation in macroscopic media. These results are, in
Sec.~\ref{sec:MCS-Cherenkov}, applied to spacelike MCS theory in order to
obtain a heuristic understanding of the expressions found in
Sec.~\ref{sec:MCS}. In Sec.~\ref{sec:modM-Cherenkov}, we give a
qualitative discussion of vacuum Cherenkov radiation in another
Lorentz-noninvariant theory with a modified-Maxwell term in the action. It
is shown that UHECRs have the potential to set tight bounds on the
``coupling constants'' of the modified-Maxwell term (details are given in
Appendix~\ref{sec:appendix-UHECRbounds}, together with a new bound based
on the already available data). In Sec.~\ref{sec:summary}, we summarize
our findings and discuss possible implications.

As to notation and conventions, we employ the Cartesian coordinates
$(x^\mu)$ $=$ $(x^0,\boldsymbol{x})$ $=$ $(c\,t,x^1,x^2,x^3)$,
the Minkowski metric $(\eta_{\mu\nu})$ $=$ $\diag(+1$,$-1$,$-1$,$-1)$,
and the totally antisymmetric Levi-Civita symbol
$\epsilon_{\mu\nu\rho\sigma}$ with normalization $\epsilon_{0123}=1$.
Indices are lowered with the Minkowski metric $\eta_{\mu\nu}$
and raised with the inverse metric $\eta^{\mu\nu}$.
In most equations, we use natural units with $c=\hbar=1$ but not always.

\section{MCS-theory Cherenkov radiation}
\label{sec:MCS}

\subsection{Spacelike MCS theory}
\label{sec:MCS-spacelike-theory}

The electromagnetic
MCS theory \cite{Carroll-etal1990,AdamKlinkhamer2001,AdamKlinkhamer2003}
has the following action:
\begin{equation}\label{eq:MCS-action}
\action_\text{MCS} = \int_{\mathbb{R}^4} \mathrm{d}^4 x\;
\Big( -\textstyle{\frac{1}{4}}\,
F_{\mu\nu}(x)\,F^{\mu\nu}(x)\,
+\,\textstyle{\frac{1}{4}}\,\, m\,\epsilon_{\mu\nu\rho\sigma}\,
\zeta^{\,\mu}\,A^\nu(x)\, F^{\rho\sigma}(x)\, \Big),
\end{equation}
with gauge field $A_\mu(x)$, Maxwell field strength
$F_{\mu\nu}(x)\equiv\partial_\mu A_\nu(x)-\partial_\nu A_\mu(x)$,
Chern--Simons (CS) mass scale $m$, and fixed normalized CS vector
$\zeta^{\,\mu}$. The background CS vector $\zeta^{\,\mu}$ can be timelike
($\zeta^{\,\mu}\zeta_{\,\mu} \equiv\zeta^{\,\mu}\,\eta_{\mu\nu}\,\zeta^{\,\nu}=1$),
null ($\zeta^{\,\mu}\zeta_{\,\mu}=0$ with $\zeta^{\,0}=1$),
or spacelike ($\zeta^{\,\mu}\zeta_{\,\mu}=-1$).

The timelike MCS theory appears to be inconsistent, that is,
the theory violates unitary or causality, or both
\cite{Carroll-etal1990,AdamKlinkhamer2001}.
In the present article, we specialize to the purely spacelike case,
\begin{subequations}\label{eq:zeta-m}
\begin{equation}\label{eq:zeta}
(\zeta^\mu)\equiv(0, \vec \zeta)\equiv(0,0,0,1),
\end{equation}
and assume the CS mass scale $m$ to be strictly nonzero
and positive,
\begin{equation}\label{eq:m}
m > 0.
\end{equation}
\end{subequations}
The condition \eqref{eq:zeta} makes the propagation of light anisotropic
and defines a class of preferred inertial frames, contradicting, thereby,
the two axioms of the theory of special relativity \cite{Einstein1905}.

The spacelike MCS theory has two photon modes with dispersion relations
\begin{equation}\label{eq:MCSdisprel}
\omega_\pm (\vec k)^2 = \abs{\vec k}^2 \pm m\,
\sqrt {\abs {\vec k}^2 \cos^2 \theta +  m^2/4} + m^2/2,
\end{equation}
where $\theta$ is the angle between wave vector $\vec k$ and
CS vector $\vec\zeta$, or, more specifically,
$\cos\theta = (\vec k \cdot \vec\zeta)/|\vec k|$.
The two polarization modes are
denoted $\plus$ and $\minus$, corresponding to the different signs in
\eqref{eq:MCSdisprel}. Further details on these polarization modes
can be found in Sec.~2 of Ref.~\cite{AdamKlinkhamer2003} and Appendix~A of
Ref.~\cite{KaufholdKlinkhamer2006}. Note, finally, that the photon theory
\eqref{eq:MCS-action}--\eqref{eq:zeta-m}
is translation invariant but not rotation invariant,
so that physical processes involving these photons
(assuming Lorentz-invariant interactions)
necessarily conserve energy-momentum but need
not conserve angular momentum.

\subsection{Vacuum Cherenkov radiation}
\label{sec:MCS-vacuum-Cherenkov-radiation}

We now add particles with electric charge $e$ and mass $M$ to the theory,
taking the usual minimal coupling to the gauge field
(i.e., replacing $\partial_\mu$ by $\partial_\mu + \ii\, e\, A_\mu$).
The action of these charged particles is assumed to be Lorentz invariant,
so that the Lorentz violation of the combined theory  resides solely
in the second term of \eqref{eq:MCS-action}. The $\minus$ photon has,
in fact, a \emph{spacelike} four-momentum, which
allows for Cherenkov radiation from any type of charged particle
with mass $M$ and three-momentum $\vec q$,
provided $\vec q \cdot \vec \zeta \neq 0$ \cite{KaufholdKlinkhamer2006}.
Hence, MCS-theory Cherenkov radiation has no threshold, in contrast
to the situation for standard Cherenkov radiation
in a nondispersive macroscopic medium
(cf. Sec.~\ref{sec:standard-Cherenkov-classical}).
For later use, we introduce the following notations:
\begin{equation}\label{eq:q-parallel-def}
\widehat{\vec q} \equiv \vec q/|\vec q|   \,,\;\;\;
q_\parallel \equiv \vec q \cdot \vec \zeta\,,\;\;\;
\vec q_\perp \equiv q_\perp\,\vec q_\perp/|\vec q_\perp|
             \equiv \vec q  - q_\parallel\, \vec \zeta\,,
\end{equation}
for the unit vector in the $\vec q$ direction and
momentum components parallel and orthogonal to the normalized CS
vector $\vec\zeta$. Remark that $q_\perp$ is defined to be nonnegative,
whereas $q_\parallel$ can have an arbitrary sign.

For the purpose of calculating the decay width,
only the $q_\parallel$ dependence is nontrivial.
We assume $\abs{q_\parallel} \gg m$ and $M\gg m$,
where $m$ is the positive CS mass scale and $M$ the mass of the charged
particle. For MCS-theory Cherenkov radiation with $\abs{q_\parallel} \sim M$,
the allowed photon momentum component is very small,
$|k_\parallel|$  running from zero to $\BigO(m)$, while,
for $\abs{q_\parallel} \gg M$, the $|k_\parallel|$
maximum is roughly equal to $\abs{q_\parallel}$.

%
\begin{figure*}[t]   
\vspace*{0cm}
\begin{center}
\includegraphics[width=4.2cm]{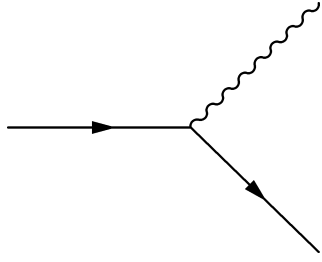}
\end{center}
\vspace*{-0.5cm}
\caption{Feynman diagram contributing to vacuum Cherenkov radiation
 from charged particles coupled to MCS photons.}
\label{fig:FDcherenkov}
\end{figure*}

The tree-level amplitude square follows from the usual Feynman
amplitude (Fig.~\ref{fig:FDcherenkov}). The result for a charged scalar is
found to be given by
\begin{equation}
|A|^2_\text{scalar} = e^2 \,\Big( 4 q^\mu q^\nu - 2  q^\mu k^\nu
 - 2 q^\nu k^\mu +  k^\mu k^\nu\Big)\,
\bar \epsilon_\mu(k) \epsilon_\nu(k),
\end{equation}
where $q^\mu$ is the initial four-momentum of the charged particle
and $k^\mu$ the four-momentum of the emitted  photon.
The result for a charged Dirac spinor,
averaged over initial spin components
and summed over final spin components, has already been calculated in
Ref.~\cite{KaufholdKlinkhamer2006} and reads
\begin{equation}
|A|^2_\text{spinor} = e^2 \,\Big( 4 q^\mu q^\nu - 2  q^\mu k^\nu -
2 q^\nu k^\mu +   k^2 \eta^{\mu\nu}\Big)\,
\bar \epsilon_\mu(k) \epsilon_\nu(k).
\end{equation}
For both amplitude squares, the polarization rule
\cite{KaufholdKlinkhamer2006} gives
\begin{equation}\label{eq:pol-sum}
\bar\epsilon_\mu (k) \, \epsilon_\nu (k) \, \mapsto \,
\frac{1}{2 k^2 + m^2\,\zeta^2}\big(-k^2\, \eta_{\mu\nu}
- m^2\,\zeta_\mu \zeta_\nu
+ \ii \,m\, \epsilon_{\mu\nu\rho\sigma}\zeta^\rho k^\sigma\big)\text.
\end{equation}
The decay widths of a particular momentum state of the scalar and
spinor particles are then defined as the phase-space integral of the
resulting amplitude squares,
\begin{equation}\label{eq:Gamma}
\Gamma_\text{scalar/spinor} \equiv
\int \mathcal{D} k \; |A|^2_\text{scalar/spinor},
\end{equation}
where $\mathcal{D} k$ is a shorthand notation for the phase-space measure;
see Ref.~\cite{KaufholdKlinkhamer2006} for details.

The decay widths for large parallel momentum components,
$\abs{q_\parallel} \gg M$, are found to be given by
\begin{equation}\label{eq:decay-scalar}
\Gamma_\text{scalar} =
\frac{1}{2}\;\alpha\, m\,
|\,\widehat{\vec q} \cdot \vec \zeta|\,
\Big(\ln ( \abs{q_\parallel}/m) + 2 \ln 2 - 1 \Big) + \cdots
\end{equation}
and
\begin{equation}\label{eq:decay-spinor}
\Gamma_\text{spinor} =
\frac{1}{2}\;\alpha\, m\,
|\,\widehat{\vec q} \cdot \vec \zeta|\,
\Big(\ln (\abs{q_\parallel}/m) +2 \ln 2 - 3/4 \Big) + \cdots,
\end{equation}
with fine-structure constant $\alpha \equiv e^2/4\pi$
(standard quantum electrodynamics of photons and electrons
has $\alpha \approx 1/137$ and $M\approx 511 \,\text{keV}$).
The exact tree-level results for the decay widths
with general three-momentum $\vec q$ are given in
Appendix~\ref{sec:appendix-decay-widths}.

Spacelike MCS theory, if physically relevant, thus predicts a
\emph{direction-dependent} lifetime of high-energy charged
particles due to vacuum Cherenkov radiation. In principle,
this nonstandard energy-loss mechanism may affect
the propagation of UHECRs.
Very possibly, even the numbers may work out, as the
following example demonstrates.

Assume the CS mass scale $m$ to be of cosmological origin
(perhaps due to new low-energy/large-distance physics)
and its numerical value to be given by
the inverse of the size of the visible universe,
$m \sim 1/L_0 \approx 1/(10^{10}\,\text{lyr})
\approx 2 \times 10^{-33}\, \text{eV}$, which is consistent
with astrophysical bounds \cite{Carroll-etal1990,Wardle-etal1997}.
Also assume primary cosmic-ray protons with energy
$E_\text{p} = 10^{19}\,\text{eV}$ (i.e., just below the
Greisen--Zatsepin--Kuzmin cutoff;
cf. Refs.~\cite{BhattacharjeeSigl1998,Stanev2004,Armengaud2005})
to have traveled over cosmological distances of the order of
$L_0 \sim 1/m$. With
$q_\parallel\equiv\vec q\cdot\vec\zeta\sim E_\text{p}\cos\theta_\text{p}$ and
$\alpha\ln (E_\text{p}/m)=\BigO(1)$,
the decay width \eqref{eq:decay-spinor} for these $10^{19}\,\text{eV}$
protons becomes $\Gamma_\text{p} \sim (m/2)|\cos\theta_\text{p}|$ and there
results a modest anisotropy for $10^{19}\,\text{eV}$ protons observed
on Earth, with
somewhat less protons from directions $\widehat{\vec q} =\pm\, \vec \zeta$
(having $|\cos\theta_\text{p}|=1$)
than from orthogonal directions $\widehat{\vec q}\perp \vec \zeta$
(having $|\cos\theta_\text{p}|=0$).
Alternatively, the lack of large-scale anisotropy
(as suggested by the current experimental data
\cite{BhattacharjeeSigl1998,Stanev2004,Armengaud2005}
and assuming the absence of strong extragalactic magnetic fields)
would place a further upper bound on the mass scale $m$ of spacelike MCS
theory at approximately $10^{-33}\,\text{eV}$
and perhaps a factor $10$ better with the complete Auger data set.
In this brief discussion, the expansion of the universe has not
been taken into account, but this can, in principle, be done to leading
order in $m$ \cite{Kostelecky2003,KantKlinkhamer2005}.

Needless to say, the example of the previous paragraph has been
given for illustrative purposes only. But the fact remains
that Lorentz-violating processes such as the one studied in the
present article could turn out to be relevant to UHECR physics.
Indeed, for cosmological applications of vacuum Cherenkov radiation,
another important quantity to calculate is the radiation rate,
which will be done in the next subsection.

\subsection{Radiated energy rate}
\label{sec:MCS-radiated-energy-rate}

The energy-momentum loss of the charged particle (scalar or spinor)
per unit of time is
equal to the photon four-momentum weighted by the amplitude square and
integrated over phase-space,
\begin{equation}\label{eq:dPmudt}
\frac{\dd P^\mu}{\dd t} \equiv \int  \mathcal{D} k\;|A|^2\,k^\mu.
\end{equation}
Making the \emph{Ansatz}
\begin{equation}\label{eq:dPmudt-coeff}
\frac{\dd P^\mu}{\dd t} =
\alpha\,m \, \big(\, K \: q^\mu + L \,m\,  \zeta^\mu \,\big),
\end{equation}
one can calculate the dimensionless coefficients $K(q_\perp,q_\parallel)$
and $L(q_\perp,q_\parallel)$ for the spacelike MCS theory
considered; see Appendix~\ref{sec:appendix-radiation-rate-coeff}.
The time component of this
last expression then gives the rate of total radiated energy,
\begin{equation}\label{eq:dWtotaldt}
\frac{\dd W}{\dd t}
\equiv
\frac{\dd P^0}{\dd t} = \alpha\,m\, K \, q^0,
\end{equation}
for $\zeta^0=0$ from \eqref{eq:zeta}.
As $t$ corresponds to the laboratory time,
the path length of the charged particle is given by $l = \beta c t$,
at least for uniform motion of the charged particle  (i.e., neglecting
radiation backreaction).

\begin{figure}
\includegraphics[width=11cm]{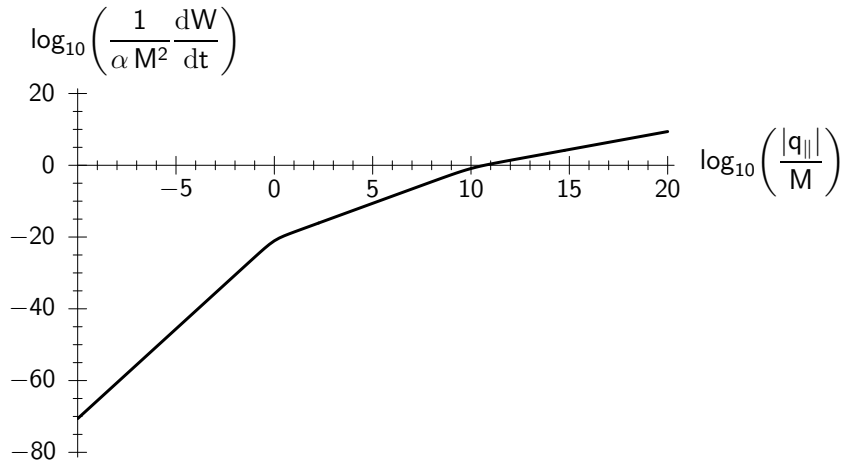}
\caption{Radiated energy rate $\dd W/\dd t$
from MCS-theory Cherenkov radiation, in units of
$\alpha \, M^2$ and as a function of $|q_\parallel|/M$,
for a charged scalar particle with mass $M$, electric charge
$e \equiv \sqrt{4\pi\alpha}$, and
parallel momentum component $q_\parallel$ as defined by \eqref{eq:q-parallel-def}.
The radiated energy rate is calculated from \eqref{eq:dWtotaldt}
and \eqref{eq:Kscalar}, for a particular choice of the CS mass scale,
$m=10^{-10} M$. The curve for a charged Dirac particle
hardly differs from the curve for a charged scalar particle shown here.}
\label{fig:rate}
\end{figure}

The radiated energy rate (Fig.~\ref{fig:rate})
has three qualitatively different domains,
again assuming $\abs{q_\parallel} \gg m$ and $M \gg m$,
while keeping $\vec q_\perp$ arbitrary.
For low momentum components compared to the particle mass,
$m \ll \abs{q_\parallel}\ll M$, the radiation rate is
essentially  the same for a scalar or spinor particle and given by
\begin{equation}\label{eq:radiated-low}
\frac{\dd W}{\dd t} = \frac{\alpha}{4}\,
\frac{m^2 \, \abs{q_\parallel}^5}{M^5}
+ \BigO(\alpha\, m^3 \abs{q_\parallel}^5/M^6).
\end{equation}
For intermediate momentum components, $m M \ll m \abs{q_\parallel}\ll M^2$,
the radiation rate is again spin independent to leading order:
\begin{equation}\label{eq:radiated-intermediate}
\frac{\dd W}{\dd t} =
\frac{\alpha}{4}\, \frac{m^2 \, \abs{q_\parallel}^2}{M^2}
+\BigO(\alpha\,m^2)\,.
\end{equation}
Our results \eqref{eq:radiated-low} and \eqref{eq:radiated-intermediate}
agree with the expression (21) obtained classically
by Lehnert and Potting \cite{LehnertPotting2004PRD70}
for a particular charge distribution and in the limit
$M/m \to \infty$.\footnote{\label{ftn:polarization}We disagree,
however, with the polarization
pattern shown in Fig.~2 of Ref.~\cite{LehnertPotting2004PRD70}.
The emitted radiation
consists solely of $\minus$ photons ($\plus$ photons are kinematically
not allowed), so that the radiation is essentially left polarized for
wave vectors $\vec k$ in the approximate hemisphere around $\vec \zeta$ and
essentially right polarized for wave vectors $\vec k$ in the approximate
hemisphere around $-\vec \zeta$, with elliptical polarizations in a
narrow band given by $|\vec k \cdot \vec \zeta| \lesssim m$.
(For details on the $\minus$ polarization mode, see, e.g., the paragraphs
below Eq.~(2.13) of Ref.~\cite{AdamKlinkhamer2003}.)
Most photons are emitted in a narrow cone around
the direction $\widehat{\vec q}$ of the charged particle
(cf. Sec.~\ref{sec:MCS-Cherenkov-angle-rate}) and have,
for generic $\widehat{\vec q}$ (that is, $|\vec q \cdot \vec \zeta|\gg m$),
the same circular polarization,
left or right depending on the sign of $\widehat{\vec q} \cdot \vec \zeta$.
With the conserved helicity of the relativistic charged particle, the
MCS Cherenkov process for generic $\widehat{\vec q}$ and
large photon momentum component $|k_\parallel| \sim |q_\parallel| \gg M$
manifestly violates angular-momentum conservation
\cite{KaufholdKlinkhamer2006}.}

For large momentum components, $m \abs{q_\parallel}\gg M^2$,
the radiated energy rate depends on the spin of the charged particle:
\begin{equation}\label{eq:radiated-scalar}
\frac{\dd W_\text{scalar}}{\dd t} = \frac{\alpha}{4}\,
m \, \abs{q_\parallel} + \BigO(\alpha\, M^2),
\end{equation}
and
\begin{equation}\label{eq:radiated-spinor}
\frac{\dd W_\text{spinor}}{\dd t} = \frac{\alpha}{3}\,
m \, \abs{q_\parallel} + \BigO(\alpha\, M^2),
\end{equation}
with only a linear dependence on the initial parallel momentum component,
compared to the quadratic dependence \eqref{eq:radiated-intermediate}
for intermediate momentum components.
The crossover from quadratic to linear behavior occurs
at a momentum component $|q_\parallel|$ of the order of $M^2/m$.
For the case of an electron, this crossover momentum takes the
following numerical value:
\begin{equation}\label{eq:crossover}
\abs{q_\parallel}_\text{\,crossover} \sim M^2\,c/m \approx
1.3 \times 10^{44} \,\text{eV}/c
\,\left(\frac{M}{511 \,\text{keV}/c^2}\right)^2
\,\left(\,\frac{2 \times 10^{-33} \,\text{eV}/c^2}{m}\,\right),
\end{equation}
for an $m$ value corresponding to the inverse of the size of
the visible universe (cf. the discussion at the end of
Sec.~\ref{sec:MCS-vacuum-Cherenkov-radiation}).
The tremendous energy appearing in \eqref{eq:crossover} is, of course,
experimentally out of reach.
But the general considerations of this article may still be relevant to
MCS theory in other applications (e.g.,
ultracold atomic systems \cite{Carusotto-etal2001}
and photonic crystals \cite{Luo-etal2001})
or to other Lorentz-violating theories such as the one discussed
in Sec.~\ref{sec:modM-Cherenkov}.

At first sight, the difference between
\eqref{eq:radiated-scalar} and \eqref{eq:radiated-spinor} is
surprising since the leading-order terms of the decay width
are the same for scalars and spinors, as shown by the logarithmic
terms in \eqref{eq:decay-scalar} and \eqref{eq:decay-spinor}.
These last terms result, in fact, from the
contribution of low-energy photons in the phase-space integral
\eqref{eq:Gamma}, where
the difference between the two amplitude squares is small.
But the radiated energy rate from \eqref{eq:dPmudt} for $\mu=0$
has an additional factor $k^0$ in the integrand and the integral gives,
for sufficiently high energies of the charged particles,
different rates for scalars and spinors.
This spin dependence at ultrahigh energies is a genuine quantum effect
and cannot be seen in the classical analysis of
Refs.~\cite{LehnertPotting2004PRL93,LehnertPotting2004PRD70}.

For completeness, we give results for two further cases.
First, consider charged  massless spinors ($M=0$), still with
relatively large parallel momentum components, $|q_\parallel| \gg m$.
It may then be of interest to calculate the
radiated energy rate of left-handed and right-handed spinors
independently. The result is
\begin{equation}
\frac{\dd W_{\text{spinor,}\;M=0}}{\dd t} = \frac{\alpha}{3}\,
m \,  \abs{q_\parallel}\big(\, 1+\chi\,\sgn(q_\parallel)\, \big)
+ \BigO(\alpha\, m^2),
\end{equation}
for spinor helicity $\chi=\pm 1/2$. In the corresponding decay
width~\eqref{eq:decay-spinor}, only the subleading term $-3/4$
inside the outer parentheses
is replaced by $-3\,\big(1-2\, \chi\, \sgn(q_\parallel)\big)/4$.

Second, consider charged  massless scalar or spinor particles ($M=0$),
now with ultralow parallel momentum components, $|q_\parallel|\ll m$.
The radiated energy rate is then given by
\begin{equation}\label{eq:zeromass-lowmomentum-rate}
\frac{\dd W_{M=0}}{\dd t}
= \frac{\alpha}{5}\, \abs{q_\parallel}^2 + \BigO(\alpha\, \abs{q_\parallel}^3/m).
\end{equation}
The radiated energy rate \eqref{eq:zeromass-lowmomentum-rate}
is not directly proportional to $m$ or $m^2$, but  its numerical value
is still very much less than $\alpha\, m^2$ because of
the stated validity domain, $|q_\parallel|\ll m$.

The rest of the article is mainly concerned with a heuristic understanding
of the results obtained in this section
and to apply that understanding to another Lorentz-violating theory.

\section{Standard Cherenkov radiation}
\label{sec:standard-Cherenkov}

\subsection{Classical process}
\label{sec:standard-Cherenkov-classical}

In this section, we recall some well-known results on standard Cherenkov
radiation \cite{Cherenkov1934,Vavilov1934,Cherenkov1937,FrankTamm1934}
and refer to three monographs~\cite{Jelley1958,Zrelov1970,Afanasiev2004}
for further details
and references (useful discussions can also be found in the textbooks
\cite{PanofskyPhillips1962,LandauLifshitz1984,Jackson1999}).
Specifically, we consider the propagation of an electrically charged
particle in an isotropic dielectric
($\epsilon(\omega) \ne 1$,  $\mu(\omega)=1$) with refractive index
$n(\omega)\equiv  c\,\abs{\vec k(\omega)}/\omega = \sqrt{\epsilon(\omega)}$,
for wave vector $\vec k$
and angular frequency $\omega$ of the electromagnetic field.
The particle has classical charge $Q \ne 0$
(Coulomb potential $V= Q/(4\pi\,r)$
in Heaviside--Lorentz units),
mass $M \geq 0$, velocity $\beta \equiv v/c \leq 1$, three-momentum $\vec q$
and energy $E=\sqrt{c^2\,\abs{\vec q}^2 + M^2\,c^4}$.
In this section, we prefer to keep $c$ and $\hbar$ explicit.

Cherenkov radiation of a particular frequency $\omega$
is emitted classically over a cone which makes an
angle $\theta_\text{C}(\omega)$ with the direction of motion of the
charged particle. The numerical value of this polar angle can be
determined by a Huygens-principle construction:
\begin{equation}\label{eq:cos-classical}
\cos\theta_\text{C}(\omega) = \frac{1}{\beta\, n(\omega)}\,,
\end{equation}
as long as $\beta\, n(\omega)\geq 1$.
The emitted radiation is \emph{linearly}
polarized with the magnetic field lying along the surface of
the cone and the electric field orthogonal to it
(angular momentum is manifestly conserved for a relativistic
charged particle; compare with the last sentence of
Footnote~\ref{ftn:polarization}).
The energy radiated per unit of time and per unit of frequency
is determined by the Frank--Tamm formula,
\begin{equation}\label{eq:radiated-classical}
\frac{\dd^2 \,\textsl{w}}{\dd t\, \dd \omega} = \beta \,\frac{Q^2}{4\pi c}\,
\sin^2\theta_\text{C}(\omega)\; \omega,
\end{equation}
where, according to \eqref{eq:cos-classical},
the factor $\sin^2\theta_\text{C}(\omega)$ can be replaced by
$1-\big(\beta\, n(\omega)\big)^{-2}$.

After integration of \eqref{eq:radiated-classical} over the allowed
frequency range, the total radiated energy rate $\dd W/\dd t$ for
$\beta=1$ is infinite classically, unless the refractive index $n(\omega)$
approaches unity fast enough for large $\omega$.
Indeed, if $n(\omega)=1$ above a cutoff frequency $\omega_\text{c}$,
the total radiated energy rate is of the order of
\begin{equation}\label{eq:dWdt-cutoff}
\frac{\dd W}{\dd t} \sim \frac{Q^2}{4\pi c}\, \omega_\text{c}^2,
\end{equation}
purely by dimensional reasons.

\subsection{Quantum effects}
\label{sec:standard-Cherenkov-quantum}

The expression \eqref{eq:cos-classical} does not take energy-momentum
conservation into account if the photon has energy $\hbar \, \omega$
and effective momentum  $\hbar\, |\vec k| =\hbar\, n(\omega)\, \omega/c$.
The correct expression for the Cherenkov angle is \cite{Ginzburg1940,Cox1944}
\begin{equation}\label{eq:cos-quantum}
\cos\theta_\text{C}(\omega) = \frac{1}{\beta \,n(\omega)} \Big( 1 +
\frac{\hbar \,\omega}{2 E}\, \big(n(\omega)^2- 1\big) \Big),
\end{equation}
as long as there is a real solution for $\theta_\text{C}(\omega)$.

For constant refractive index $n$, \eqref{eq:cos-quantum} gives a maximum
photon energy
\begin{equation}
\hbar\,\omega_\text{max} = 2\, E\; \frac{\beta\, n - 1}{n^2-1}.
\end{equation}
The quantum modification \eqref{eq:cos-quantum} makes the Cherenkov
angle smaller than the classical value (as long as $\beta\, n>1$) and
the Cherenkov cone shrinks to the forward direction
as $\omega \to \omega_\text{max}$.
The total radiated energy rate is now given by
\begin{equation}\label{eq:dWdt-quadratic}
\frac{\dd W}{\dd t} \sim \frac{Q^2}{4\pi c}\,E^2/\hbar^2,
\end{equation}
up to factors of order unity. Expression \eqref{eq:dWdt-quadratic},
compared to \eqref{eq:dWdt-cutoff}, makes clear that
quantum theory ($\hbar \ne 0$) renders the radiated energy rate finite
by providing a cutoff on the frequency of the emitted radiation,
even for the case of a frequency-independent refractive index.

\subsection{Model}
\label{sec:standard-Cherenkov-model}
\vspace*{-.25\baselineskip}

Consider, next,  a refractive index which is assumed to behave as follows:
\begin{equation}\label{eq:n-model}
n(\omega)\,\big|^\text{model} = 1 + \frac{\omega_0}{2\, \omega},
\end{equation}
for $\omega>\omega_0$, with fixed angular frequency $\omega_0$.
The assumed behavior of \eqref{eq:n-model} certainly
corresponds to ``anomalous dispersion,'' but the functional dependence
on $\omega$ is very different from that of standard macroscopic media
with $n(\omega) \sim 1-\omega_p^2/\omega^2$ for $\omega \to \infty$;
cf. Sec.~7.5 of Ref.~\cite{Jackson1999}.
The precise form of \eqref{eq:n-model} is, in fact, chosen
for comparison to MCS theory, as will become clear in the next section.

For the special behavior \eqref{eq:n-model} of the refractive index,
the quantum correction term in \eqref{eq:cos-quantum} turns out to be
essentially frequency independent for $\omega \gg \omega_0$,
\begin{equation}\label{eq:cos-model}
\cos\theta_\text{C}(\omega) \,\big|^\text{model} =
\frac{1}{\beta\, n(\omega)} \bigg( 1 + \frac{\hbar \,\omega_0}{2\,E}
\big(1 + \BigO(\omega_0/\omega)\big) \bigg),
\end{equation}
making the classical Cherenkov angle a good approximation for large
particle energy $E\gg \hbar\,\omega_0$.
For the model considered, the cutoff frequency is given
by $\omega_\text{max} = E/\hbar$ and the total radiated energy
rate is finite.\footnote{Classically, the radiation output in a medium
with refractive
index \eqref{eq:n-model} would be proportional to $\omega_0^2$.
This output would, however, be finite only for
$\beta<1$, because $n(\omega)$ does not approach unity fast enough
for $\omega \to \infty$.}
Making the replacements $\beta =c\, |\vec q|/E$
and $E=\sqrt{c^2\,|\vec q|^2+M^2\,c^4}$,
one obtains the following high-energy behavior:
\begin{equation}\label{eq:radiated-model}
\frac{\dd W}{\dd t}\,\bigg|^\text{model} =
\frac{1}{2}\,\frac{Q^2}{4\pi \hbar c}\,
\omega_0 \, E + \cdots,
\end{equation}
where the ellipsis contains terms (involving logarithms)
which are, at ultrahigh energies, small compared to the term shown.
Note that the radiated energy rate \eqref{eq:radiated-model}
only grows linearly with $E$, compared to the quadratic behavior
\eqref{eq:dWdt-quadratic} for the case of constant refractive index.

\section{Heuristics of MCS-theory Cherenkov radiation}
\label{sec:MCS-Cherenkov}

\subsection{Refractive index}
\label{sec:MCS-Cherenkov-n}

High-energy $\minus$ photons of MCS theory
moving in a generic direction $\widehat{\vec k}$
have, according to \eqref{eq:MCSdisprel}, a refractive index given by
\begin{equation}\label{eq:n-minus-MCS}
n_\minus(\vec{k})
\equiv \abs{\vec k} / \omega_{-}(\vec k)
= 1+ \frac{m \, \abs{\cos\theta}}{2 \abs{\vec k}}+\BigO(m^2/\abs{\vec k}^2),
\end{equation}
for $|\cos\theta| \equiv |\widehat{\vec k} \cdot \vec \zeta| \gg m/|\vec k|$
and $c=\hbar=1$. This is, in fact, the refractive index
relevant to MCS-theory Cherenkov radiation from a highly energetic charged
particle moving in a generic direction, where most
radiation energy is carried away by photons with
large parallel momentum components,
$|k_\parallel|\equiv |\vec k \cdot \vec \zeta| \gg m$.
Dropping the suffix $\minus$ on $n$, the refractive index
\eqref{eq:n-minus-MCS} can be written as
\begin{equation}\label{eq:n-model-MCS}
n\big(\omega,\widehat{\boldsymbol{k}}\big)
\,\big|^\text{model}_\text{MCS}
= 1 + \frac{m\,|\cos\theta|}{2\omega} + \BigO(m^2/\omega^2),
\end{equation}
in order to connect to the particular model discussed
in Sec.~\ref{sec:standard-Cherenkov-model}.

\subsection{Cherenkov angle and radiated energy rate}
\label{sec:MCS-Cherenkov-angle-rate}

Setting $\hbar\,\omega_0 \sim m c^2$ in the model result~\eqref{eq:cos-model}
shows that, for high energies $E$ of the charged particle,
the quantum correction to the Cherenkov angle goes to zero as $m c^2/E$.
This behavior is quite different from that of
standard Cherenkov radiation in a nondispersive medium, as given by
\eqref{eq:cos-quantum} for $\hbar\,\omega \sim E/2$ and constant $n$.
The ``good'' high-energy behavior of MCS theory is perhaps not entirely
surprising as the nonstandard term in the action \eqref{eq:MCS-action}
is super-renormalizable.

Specifically, for large energy $E \gg M \gg m$
and generic direction $\widehat{\vec q}$ of the charged particle
($\widehat{\vec q} \cdot \vec \zeta \ne 0$),
the MCS-theory Cherenkov radiation is emitted in a pencil beam around
the $\widehat{\vec q}$ direction with an angular dimension of the order of
\begin{equation}\label{eq:thetaC-MCS}
2\,\theta_\text{C}\big(\omega,\widehat{\boldsymbol{q}}\big)
\,\big|^\text{model}_\text{MCS} \sim
2\,\sqrt{|\widehat{\vec q} \cdot \vec \zeta|\,m c^2/(\hbar\,\omega)}\;
\left(\, 1+ \BigO\big(Mc^2/E,mc^2/(\hbar\,\omega)\big)\,\right),
\end{equation}
for frequencies $\omega$ up to $\omega_\text{max} = E/\hbar$
and with $\hbar$ and $c$ temporarily reinstated.\footnote{The mass scale
$m$ enters the action \eqref{eq:MCS-action} through the
combination $m\,c/\hbar \equiv 1/\ell$,
in terms of a fundamental length scale $\ell$.
In this way, the leading term of \eqref{eq:thetaC-MCS}
can be written as
$2\,\sqrt{|\widehat{\vec q} \cdot \vec \zeta|\,c/(\ell\,\omega)}$,
without explicit Planck constant $\hbar$.}
The factor $|\widehat{\vec q} \cdot \vec \zeta|$
under the square root of \eqref{eq:thetaC-MCS}
corresponds to the absolute value of the cosine of the angle
between the charged particle direction
$\widehat{\vec q}$ and the fixed CS direction $\vec \zeta$,
which, in turn, traces back to the cosine factor in the refractive index
\eqref{eq:n-model-MCS} of the individual photons.

We have also calculated the radiated energy rate $\dd W/\dd t$ from
the model result~\eqref{eq:radiated-model}, replacing $Q^2/(4\pi \hbar c)$
by $\alpha$ as defined below \eqref{eq:decay-spinor}
and inserting an overall factor of $1/2$.
This extra factor $1/2$ for the radiated energy rate is due
to the fact that only one photon polarization ($\minus$) contributes in
MCS theory, the $\plus$ photon having a timelike four-momentum.
For intermediate momentum components, $m M < m \abs{q_\parallel}< M^2$,
the adapted model result \eqref{eq:radiated-model},
taking into account the terms not shown explicitly,
is numerically in good agreement with the
original expression \eqref{eq:radiated-intermediate}.
For ultrahigh momentum components, $m \abs{q_\parallel} \gg M^2$,
the adapted model result~\eqref{eq:radiated-model}, with $\omega_0$ replaced by
$m|\cos\theta|$, gives immediately
\begin{equation}
\frac{\dd W}{\dd t}\,\bigg|^\text{model}_\text{MCS}
 = \frac\alpha 4\: m\, \abs{q_\parallel}
+ \cdots\,,
\end{equation}
which agrees with the scalar radiation rate
\eqref{eq:radiated-scalar} calculated directly.
The explanation for the slightly different spinor radiation
rate \eqref{eq:radiated-spinor} will be given in the next subsection.

\subsection{Spin effects}
\label{sec:MCS-Cherenkov-spin}

A charged particle with spin has a different interaction with the
photon as a charged particle without spin.
The Cherenkov radiation of a charged Dirac particle
receives, therefore, an extra contribution
(e.g., from the magnetic dipole moment)
compared to the case of a charged scalar particle.
From Eq.~(2.39) of Ref.~\cite{Jelley1958} or Eq.~(280) of
Ref.~\cite{Zrelov1970}, we obtain the following extra contribution
for a particle of spin $1/2\,$:
\begin{equation}\label{eq:spin}
\Delta \bigg(\frac{\dd W_\text{spinor}}{\dd t}\bigg)
=  \frac{\alpha}{\beta}
\,\int_0^{\omega_\text{max}} \id \omega \; \hbar\,\omega\;
\bigg(\, \frac{\hbar^2 \omega^2}{4\, E^2}
\,\big(n(\omega)^2-1 \,\big)\,\bigg).
\end{equation}
Using the refractive index \eqref{eq:n-model-MCS}
and $\omega_\text{max}=E \gg m$
(again setting $c=\hbar=1$), this expression gives a
contribution with a linear momentum dependence,
\begin{equation}
\Delta \bigg(\frac{\dd W_{\text{spinor}}}{\dd t} \bigg)
\,\bigg|^\text{model}_\text{MCS} =
\frac{\alpha}{12} \,m \, \abs{q_\parallel} + \cdots,
\end{equation}
which explains the difference
between \eqref{eq:radiated-scalar} and \eqref{eq:radiated-spinor}.

The crossover between the spin-independent behavior of
\eqref{eq:radiated-intermediate} and the spin-dependent behavior of
\eqref{eq:radiated-scalar}--\eqref{eq:radiated-spinor} occurs
at a momentum component $|q_\parallel|$ of order $M^2/m$, which has already
been discussed in Sec.~\ref{sec:MCS-radiated-energy-rate}.

\section{Modified-Maxwell-theory Cherenkov radiation}
\label{sec:modM-Cherenkov}

The main focus of this article has been on explicit
calculations of vacuum Cherenkov radiation in
spacelike MCS theory coupled to Lorentz-invariant
charged particles. In this section, we give a qualitative
discussion of vacuum Cherenkov radiation in
the only other possible gauge-invariant renormalizable theory for photons
with Lorentz violation, namely, the so-called modified-Maxwell theory.

The action for modified-Maxwell theory can be written as follows
\cite{ChadhaNielsen1982,ColladayKostelecky1998}:
\begin{equation}\label{eq:modM-action}
\action_\text{modM} =
\int_{\mathbb{R}^4} \id^4 x \;
\Big( -\textstyle{\frac{1}{4}}\,
\big(
\eta^{\mu\rho}\eta^{\nu\sigma}
+\kappa^{\mu\nu\rho\sigma}\big) \,
F_{\mu\nu}(x)\,F_{\rho\sigma}(x)
\Big)\,,
\end{equation}
for a real dimensionless background tensor $\kappa^{\mu\nu\rho\sigma}$
having the same symmetries as the Riemann curvature tensor and
a double trace condition $\kappa^{\mu\nu}_{\phantom{\mu\nu}\mu\nu}=0$
(so that there are $20-1=19$ independent components).
All components of the $\kappa$--tensor in \eqref{eq:modM-action}
are assumed to be very small, $|\kappa^{\mu\nu\rho\sigma}|\ll 1$.
Remark that the $\kappa FF$ term in
\eqref{eq:modM-action} is CPT even, whereas
the $mAF$ term in  \eqref{eq:MCS-action} is CPT odd.

Vacuum Cherenkov radiation for standard electrodynamics
with the modified photonic action \eqref{eq:modM-action}
has been studied classically by
Altschul \cite{Altschul2007PRL98,Altschul2007PRD75}.
Here, we can already make some general remarks on quantum effects,
keeping $\hbar$ and $c$ explicit for the remainder of this section.

The nonstandard term in the action \eqref{eq:modM-action} is scale
invariant, just as the standard term, and the modified photon dispersion
relation is given by
\begin{equation}\label{eq:other-disp-rel}
\omega(\vec k)^2 = c^2\,|\vec k|^2 \,
\big(1 - \Theta(\widehat{\vec k})\big)\,\text,
\end{equation}
with $\widehat{\vec k}$ the unit vector in the direction of $\vec k$
and $\Theta$ a particular function of $\widehat{\vec k}$,
the components of the $\kappa$--tensor being considered fixed.
Hence, the refractive index $n=1/\sqrt{1-\Theta}$
depends on direction, not frequency.

However, according to \eqref{eq:cos-quantum}
with $n(\omega)$ replaced by $n(\widehat{\vec k})$,
quantum effects make the Cherenkov angle frequency dependent
by an additional term proportional to the ratio of photon
energy $\hbar\,\omega$ and particle energy $E$. Assuming
a massless charged particle ($M=0$, $\beta\equiv v/c=1$) and refractive
index $n=1+\delta n$ with $\delta n= \delta n(\widehat{\vec k})\geq 0$,
we have for the Cherenkov relation \eqref{eq:cos-quantum}:
\begin{equation}\label{eq:other-costhetaC}
\cos\theta_\text{C}= 1 - \delta n \,(1 - \hbar \,\omega/E) +
\BigO(\delta n^2),
\end{equation}
and for the corresponding factor entering the
differential radiated energy rate \eqref{eq:radiated-classical}:
\begin{equation}\label{eq:other-sin2thetaC}
\sin^2\theta_\text{C}= 2\, \delta n \,( 1- \hbar \,\omega /E) +
\BigO(\delta n^2),
\end{equation}
so that anisotropy effects from $\delta n$ occur already at zeroth order
in $ \hbar \,\omega /E$ and explicit quantum effects at first order.
Spin effects are of higher order in $\omega$
but of the same order in $\delta n$, according to \eqref{eq:spin}.

Under the conditions stated, one expects for the generic radiated energy
rate of a particle with electric charge $e \equiv \sqrt{4\pi\alpha}$,
mass $M\geq 0$, momentum $\boldsymbol{q}$,
and energy $E \sim c\,|\boldsymbol{q}|$:
\begin{equation}\label{eq:other-dWdt}
\frac{\dd W_\text{modM}}{\dd t}
= \alpha\, (\kappa q q)\, c^2/\hbar \sim
\alpha\, \big(\xi_0 + \xi_1(\widehat{\vec q}) \big)\, E^2 / \hbar\,
\Big|_{E \gg E_\text{thresh}}\;,
\end{equation}
in terms of the highly symbolic notation $(\kappa q q)$ for the appropriate
contractions of the \mbox{$\kappa$--tensor} with two \mbox{$q$--vectors}
and nonnegative coefficients $\xi_0$ and $\xi_1$, the
latter coefficient having a nontrivial direction dependence.
The asymptotic behavior shown in \eqref{eq:other-dWdt}
holds only for particle energies $E$ well above the (direction-dependent)
Cherenkov threshold, which has an order of magnitude given by
\begin{equation}\label{eq:other-Ethresh}
E_\text{thresh} \sim M c^2/\sqrt{\kappa},
\end{equation}
for an appropriate scale $\kappa$
obtained from the $\kappa^{\mu\nu\rho\sigma}$ components
($\kappa$ is effectively set to zero if Cherenkov radiation is not allowed).

Our estimate for the threshold energy agrees with the result obtained by
Altschul \cite{Altschul2007PRL98}, as given by his Eq.~(4) for a subset of
$\kappa$--components (see below). His treatment of the radiation rate,
however, is purely classical, as it neglects quantum effects on the
Cherenkov angle and the differential radiation rate. He, then, introduces
an energy cutoff $\Lambda$ (possibly related to  ``new physics'' which may
or may not be required by causality) to make the total radiated energy
rate finite, $\dd  W/\dd t \sim \alpha\, \kappa\, \Lambda^2 / \hbar$, as
given by his Eq.~(7). But, as discussed in our
Sec.~\ref{sec:standard-Cherenkov-quantum}, such a cutoff is already
provided in the quantum theory by the energy $E$ of the particle. With
this cutoff $E$, the radiation rate above threshold is really given by
$\dd W/\dd t  \sim \alpha\,\kappa \, E^2 / \hbar$, in agreement with
\eqref{eq:other-dWdt} above. The difference between the asymptotic energy
behaviors of the radiation rates \eqref{eq:radiated-spinor} and
\eqref{eq:other-dWdt} traces back to the fact that the CS term in
\eqref{eq:MCS-action} has a single derivative and the $\kappa F F$ term in
\eqref{eq:modM-action} has two. Once more, the total radiated energy rate
for both Lorentz-violating theories is finite because of the frequency
cutoff from standard quantum mechanics.

Possible signatures of the Lorentz-violating action \eqref{eq:modM-action}
and the corresponding radiation rate \eqref{eq:other-dWdt}
include nonstandard propagation effects for UHECRs,
similar to the MCS effects discussed in the last three paragraphs
of Sec.~\ref{sec:MCS-vacuum-Cherenkov-radiation}. In order to be specific,
let us follow Altschul \cite{Altschul2007PRL98} by
keeping only nine of the nineteen independent
``coupling constants'' from $\kappa^{\mu\nu\rho\sigma}$
in \eqref{eq:modM-action},
namely, those which do \emph{not} lead to birefringence.
Precisely these coupling constants, for flat spacetime denoted
$\widetilde{\kappa}^{\mu\nu}\equiv \kappa^{\rho\mu\sigma\nu}\,\eta_{\rho\sigma}$
(symmetric and traceless in $\mu,\nu$),
are only constrained at the $10^{-16}$ level or
worse \cite{KosteleckyMewes2002,Carone-etal2006,Stanwix-etal2006}.
Now, taking the most energetic UHECR event known today
\cite{Bird-etal1995,Risse-etal2004}  to correspond to a primary proton
with energy $E_\text{p}  \approx 3 \times 10^{11} \, \text{GeV}\,$
and restmass $M_\text{p} \approx 0.938\, \text{GeV}/c^2$,
the Cherenkov threshold condition,
$M_\text{p}\,c^2/\sqrt{\widetilde{\kappa}}
\gtrsim E_\text{p}$, gives the following upper bound on the magnitude
of generic $\widetilde{\kappa}^{\mu\nu}$ components:
$\widetilde{\kappa} \lesssim 10^{-23}$.
(A similar bound at the $10^{-23}$ level has been
derived kinematically by Coleman and Glashow \cite{ColemanGlashow1997}.)
Further details on our bound for generic $\widetilde{\kappa}^{\mu\nu}$
and similar bounds for special (nongeneric) $\widetilde{\kappa}^{\mu\nu}$
are given in Appendix~\ref{sec:appendix-UHECRbounds}.

\section{Summary}
\label{sec:summary}

In Secs.~\ref{sec:MCS}, \ref{sec:standard-Cherenkov},
and \ref{sec:MCS-Cherenkov} of this article,
we have arrived at a detailed understanding
of vacuum Cherenkov radiation in spacelike
MCS theory \eqref{eq:MCS-action}--\eqref{eq:zeta-m}
coupled to Lorentz-invariant charged particles with spin $0$ or $1/2$.

Remarkably, quantum corrections to the Cherenkov angle $\theta_\text{C}$
and the amplitude square are suppressed because the refractive
index $n$, for large photon momenta $\abs{\vec k}$,
behaves as $1+ \BigO(\,m/\abs{\vec k}\,)$,
with $m$ the mass scale of the MCS theory.
Quantum effects enter mainly by the condition on the maximum radiated
photon energy and the resulting total radiated energy rate is, for
ultrahigh particle energy $E$, proportional to $m E$.

In addition, the effects of the charged particle's spin are not
negligible and change the coefficient of the leading term of the
radiation rate.
Since the radiated MCS photons are typically circularly polarized,
effects of angular momentum nonconservation play a role.
But, like quantum effects for the amplitude square, further
spin effects are suppressed at ultrahigh energies.

Vacuum Cherenkov radiation is quite different
in modified-Maxwell theory \eqref{eq:modM-action},
as shown in Sec.~\ref{sec:modM-Cherenkov}.
The corresponding radiation
rate has not been calculated exactly but its generic asymptotic behavior
has been established and was given in \eqref{eq:other-dWdt}.
Contrary to the case of MCS-theory Cherenkov radiation,
modified-Maxwell-theory Cherenkov radiation does have an energy threshold,
which allows for UHECR bounds on certain combinations of the nineteen
``coupling constants'' from $\kappa^{\mu\nu\rho\sigma}$
in the modified-Maxwell action \eqref{eq:modM-action}.

As discussed in Appendix~\ref{sec:appendix-UHECRbounds}, future UHECR bounds
on the nine nonbirefringent coupling constants from $\kappa^{\mu\nu\rho\sigma}$
can be expected at the $10^{-23}$ level, at least, for primary protons.
At this moment, we only have a bound at the $10^{-23}$ level
for a special choice of coupling constants (the spatially isotropic
case 2 in Appendix~\ref{sec:appendix-UHECRbounds}).
Together with existing bounds
on the ten remaining birefringent coupling constants
from $\kappa^{\mu\nu\rho\sigma}$
at the $10^{-32}$ level \cite{KosteleckyMewes2002},
these expected UHECR bounds at the $10^{-23}$ level may suggest
that the CPT--even Lorentz-violating $\kappa F F$ term
in \eqref{eq:modM-action}
is effectively absent.\footnote{The same can perhaps not be said
of the CPT--odd $m A F$ term in \eqref{eq:MCS-action},
for which there exists at least one physical mechanism
\cite{Klinkhamer2000,KlinkhamerSchimmel2002,Klinkhamer2005} that
naturally gives small values for the mass scale $m$, namely,
proportional to the inverse of the size of the universe.}
If true, this constitutes one more hint
(see, e.g., Refs.~\cite{Collins-etal2004,BernadotteKlinkhamer2007}
for other hints) in support of the fundamental role of Lorentz invariance
at the high-energy/small-distance frontier of physics.

\section*{ACKNOWLEDGMENTS}
\noindent FRK thanks E. Armengaud and M. Risse for helpful discussions
on UHECRs. The work of CK is supported by the Deutsche
For\-schungs\-ge\-mein\-schaft through the Gra\-du\-ier\-ten\-kol\-leg
``High Energy Physics and Particle Astrophysics'' (GRK742).

\begin{appendix}
\section{MCS decay widths}
\label{sec:appendix-decay-widths}

In this Appendix, analytic tree-level results for the decay widths
from MCS Cherenkov radiation are presented.
Considered are charged scalar and spinor particles with
electric charge $e$, mass $M$, and generic three-momentum
$\vec q \equiv q_\parallel \:\vec \zeta + \vec q_\perp$.
The mass-shell condition of the charged particle
is given by the standard Lorentz-invariant expression,
$E^2= q_\perp^2+q_\parallel^2+M^2$.
The theory considered is standard quantum electrodynamics
with an additional spacelike Chern--Simons (CS)
term in the photonic action, where $m$ denotes the Lorentz-violating
CS mass scale and $\vec \zeta$ the normalized CS vector.
See Secs.~\ref{sec:MCS-spacelike-theory}
and \ref{sec:MCS-vacuum-Cherenkov-radiation} for details.
All calculations of this article have been performed
with \textsc{mathematica 5.0} \cite{Wolfram99}.

The different terms of the decay widths will be ordered as follows:
first, an inverse-hyperbolic term and, then, terms with descending
powers of $m$. With this ordering and natural units ($c=\hbar=1$),
the scalar decay width is found to be given by
\begin{align}\hspace*{-1cm}
\Gamma_\text{scalar}(q_\perp,q_\parallel) = &
\frac{\alpha\,m}
{16 \, E \,\big(q_\parallel^2 + M^2\big)^{1/2}}\, \bigg(
\Big( 4\big( M^2 +2 q_\parallel^2\big)-m^2\Big)\;\arcsinh(2\kmax/m)
\nonumber \\
& + 2 \big(2 \absqp + \kmax \big)\, m
- 4 \absqp \, \sqrt{m^2 + 4 \kmax^2}
-8\,M^2\,\kmax/m
\bigg),
\end{align}
and the spinor decay width by
\begin{align}\hspace*{-1cm}
\Gamma_\text{spinor}(q_\perp,q_\parallel) = &
\frac{\alpha\,m}
{16 \, E \, \big(q_\parallel^2 + M^2\big)^{1/2}}\, \bigg(
\Big(4\big( M^2 +2 q_\parallel^2\big)+m^{2}/2 \Big)\;\arcsinh(2\kmax/m)
\nonumber \\
& + 2 \big(2 \absqp - \kmax\big)\, m
- \big(4 \absqp - \kmax\big)\,\sqrt{m^2 + 4 \kmax^2}
-8\,M^2\,\kmax/m
\bigg),
\end{align}
in terms of the fine-structure constant  $\alpha\equiv e^2/(4\pi)$
and the maximum photon momentum component $\kmax$ defined by
\begin{equation}\label{eq:kmax}
\hspace*{-1cm}
k_{\text{max}}(q_\parallel) \equiv
\frac{2 m  |q_\parallel|\,\left(m +2\, \sqrt{q_\parallel^2+M^2}\,\right)}
     {m^2+4M^2+4 m\, \sqrt{q_\parallel^2+M^2}\,} \geq 0\;.
\end{equation}
The spinor decay width has already been given in
Ref.~\cite{KaufholdKlinkhamer2006} but has been included here
for comparison with the scalar result.

\section{MCS radiation rate coefficients}
\label{sec:appendix-radiation-rate-coeff}

In this Appendix, analytic tree-level results for the coefficients
$K$ and $L$ of the MCS Cherenkov
energy-momentum-loss rate \eqref{eq:dPmudt-coeff} are presented,
the first of which completely determines the radiated energy
rate \eqref{eq:dWtotaldt} of the charged particle considered
(scalar or spinor).
See Appendix~\ref{sec:appendix-decay-widths} for further details.

The radiation rate coefficients for a charged scalar particle are given by
\begin{widetext}
\begin{subequations}
\begin{align}\hspace*{-1cm}
K_\text{scalar}(q_\perp,q_\parallel) =&
\frac{1}{128 \, E \, \big(q_\parallel^2+M^2\big)^{3/2}}\, \bigg(
3 \Big(4 \big(M^2+2 q_\parallel^2\big)-m^2\Big)\,m^2\,\arcsinh(2 \abskmax/m)
\nonumber \\
& + \big( 12 \absqp +8 \kmax\big) \, m^3
\nonumber \\
&
- 2\sqrt{m^2+4 \kmax^2}\, \big(6 \absqp+\abskmax \big) \,m^2 \nonumber \\
& - 8\Big(2 M^2 \big(\absqp+2\kmax \big) + \absqp \big(4 q_\parallel^2
+ 4 \absqp \kmax -3 \kmax^2\big) \Big) \, m \nonumber\\
& +8 \sqrt{m^2+4 \kmax^2}\,\Big(M^2 \big(2 \absqp+\abskmax \big)
+2 q_\parallel^2\, \big(2 \absqp-\abskmax\big)\Big) \nonumber \\
& - 32 M^2 \absqp\, \kmax^2  / m
\bigg),
\label{eq:Kscalar}
\end{align}
\begin{align}
\hspace*{-1cm}
L_\text{scalar}(q_\perp,q_\parallel)  =&
\frac{\sgn(q_\parallel)}{128\,E\,\big(q_\parallel^2+M^2\big)^{3/2}}\,\bigg(
\!\!- \absqp \Big(4 \big(M^2+4 q_\parallel^2\big) - 3 m^2\Big)\,
m\,\arcsinh(2\abskmax/m)
\nonumber\\
& -4 \Big(\! -M^2 + 2 \absqp \big(\absqp + \kmax\big)\Big)\, m^2\nonumber\\
&+2\sqrt{m^2+4\kmax^2}\,\Big(\! -2 M^2 + \absqp \big(4 \absqp+\abskmax \big)
\Big)\,m \nonumber\\
& -8 \Big(2 M^4 + M^2 \big(4 q_\parallel^2 - 4 \absqp \kmax- \kmax^2 \big) + 2
q_\parallel^2 \,\kmax \big(-2 \absqp+\kmax\big) \Big) \nonumber \\
& +8 M^2\sqrt{m^2+4\kmax^2}\,\Big(2 M^2+\absqp \big(4 \absqp-3
\abskmax\big)\Big)/m \nonumber \\
& - 32 M^4 \kmax^2 /m^2
\bigg),
\label{eq:Lscalar}
\end{align}
\end{subequations}
and those for a  charged spin $1/2$ particle by
\begin{subequations}
\begin{align}\hspace*{-1cm}
K_\text{spinor}(q_\perp,q_\parallel) =&
\frac{1}{192 \, E \, \big(q_\parallel^2+M^2\big)^{3/2}}\, \bigg(
3 \Big(6 \big(M^2+2 q_\parallel^2\big)+m^2\Big)\,m^2\,\arcsinh(2 \abskmax/m)
\nonumber \\
&
+ \big(10 \absqp -12 \kmax\big) \, m^3
\nonumber \\
&
-2 \sqrt{m^2+4 \kmax^2}\,  \big( 5 \absqp -3 \abskmax\big) \, m^2
\nonumber \\
& -4 \Big(6 M^2 \big( \absqp+ 2 \kmax \big) + 12\absqp^3
+ 12 q_\parallel^2 \kmax  - 3\absqp \kmax^2 + 2 \kmax^3 \Big) \, m \nonumber\\
& +4 \sqrt{m^2+4 \kmax^2}\,\Big(3 M^2 \big(2 \absqp+\abskmax\big)
+2 \absqp \big(6 q_\parallel^2-3 \absqp \abskmax + \kmax^2 \big)\Big)
\nonumber \\
& -48 M^2 \absqp\, \kmax^2 / m
\bigg),
\label{eq:Kspinor}
\end{align}
\begin{align}
\hspace*{-1cm}
L_\text{spinor}(q_\perp,q_\parallel) =&
\frac{\sgn(q_\parallel)}{192\,E\,\big(q_\parallel^2+M^2\big)^{3/2}}\,\bigg(
\!\!-3 \absqp\Big(2\big(M^2+4 q_\parallel^2\big)+m^2\Big)\,
m\,\arcsinh(2\abskmax/m)
\nonumber \\
& -2 \big(M^2 + 6 q_\parallel^2 - 6 \absqp \kmax\big) \, m^2 \nonumber \\
& +2 \sqrt{m^2+4\kmax^2}\,\Big(M^2+3\absqp\big(2 \absqp-\abskmax\big)
\Big)\,m
\nonumber\\
& -4 \Big(6 M^4 + 3 M^2 \big(4\absqp^2  - 4 \kmax\absqp+\kmax^2\big)
\nonumber\\
&
\quad\quad - 2 \absqp\kmax\big(6 q_\parallel^2 - 3 \absqp \kmax +\kmax^2 \big)\Big)
\nonumber \\
& +4 M^2 \sqrt{m^2+4 \kmax^2}\,  \Big(6 M^2+12 q_\parallel^2
-9 \abskmax \absqp+2 \kmax^2\Big)/m \nonumber \\
& -48 M^4 \kmax^2/m^2
\bigg),
\label{eq:Lspinor}
\end{align}
\end{subequations}
\end{widetext}
with $E \equiv q^0 = \sqrt{q_\perp^2+q_\parallel^2+M^2}$
the energy of the charged particle (from the initial state factor)
and $\kmax$ the maximum photon momentum component defined by
\eqref{eq:kmax}.

In closing, we remark that the factor $E$ in the denominator
of \eqref{eq:Kscalar} or \eqref{eq:Kspinor} cancels out in the
corresponding radiated energy rate \eqref{eq:dWtotaldt},
so that this rate only depends on the magnitude of the
parallel momentum component, $|q_\parallel|$.

\section{UHECR bounds on nonbirefringent modified-Maxwell theory}
\label{sec:appendix-UHECRbounds}

In this Appendix, vacuum Cherenkov bounds are discussed for certain
``coupling constants'' of modified-Maxwell theory
\eqref{eq:modM-action}
over flat Minkowski spacetime with a metric $\eta_{\mu\nu}$
as defined in Sec.~\ref{sec:introduction}.
Specifically, the following \emph{Ansatz} for $\kappa^{\mu\nu\rho\sigma}$
is considered \cite{Altschul2007PRL98}:
\begin{equation}\label{eq:widetilde-kappa-mu-nu}
\kappa^{\mu\nu\rho\sigma} =
\textstyle{\frac{1}{2}} \big(\,
\eta^{\mu\rho}\,\widetilde{\kappa}^{\nu\sigma} -
\eta^{\mu\sigma}\,\widetilde{\kappa}^{\nu\rho} -
\eta^{\nu\rho}\,\widetilde{\kappa}^{\mu\sigma} +
\eta^{\nu\sigma}\,\widetilde{\kappa}^{\mu\rho}
\,\big) ,
\end{equation}
in terms of the nine components of a
symmetric and traceless matrix $\widetilde{\kappa}^{\alpha\beta}$.
The corresponding modified-Maxwell theory has no birefringence.

It is possible to derive an upper
bound on a combination of the coupling constants
$\widetilde{\kappa}^{\mu\nu}$ from the observation of a UHECR
event with a primary proton moving in the direction $\widehat{\vec q}$
and having an ultrarelativistic energy $E_\text{p} \gg M_\text{p}\,c^2$.
Very briefly, the argument runs as follows \cite{Beall1970,ColemanGlashow1997}:
an ultra-high-energy cosmic proton
can arrive on Earth only if it does not lose energy
by vacuum Cherenkov radiation, which requires the proton energy $E_\text{p}$
to be close to or below threshold, $E_\text{p} \lesssim E_\text{thresh}$.
Using the explicit result for the threshold energy \cite{Altschul2007PRL98},
this condition can be written as the following upper bound:
\begin{equation}\label{eq:generalbound}
R\big( \widetilde{\kappa}_{ij}\,\widehat{\vec q}_{i}\,\widehat{\vec q}_{j}+
       2\,\widetilde{\kappa}_{0j}\, \widehat{\vec q}_{j}+
       \widetilde{\kappa}_{00}
       \big)
\lesssim
\big(M_\text{p}\,c^2/E_\text{p} \big)^2
=
10^{-20} \,\left(\frac{10^{10} \, \text{GeV}}{E_\text{p}}\right)^2
\,\left(\frac{M_\text{p}}{\text{GeV}/c^2}\right)^2 ,
\end{equation}
with the ramp function $R(x) \equiv x\,\theta(x)$,
defined in terms of the step function $\theta(x)=1$ for $x \geq 0$ and
$\theta(x)=0$ for $x < 0$. For a negative (spacelike) argument of the ramp
function on the left-hand-side of \eqref{eq:generalbound}, the phase velocity
of light in the specified direction is larger than $c$
and vacuum Cherenkov radiation is impossible
(the maximum attainable velocity of the charged particle
being equal to $c$); see below for further comments.

In order to simplify the discussion, it will be
assumed that many UHECRs with energy of
$10 \,\text{EeV}= 10^{10}\,\text{GeV}= 10^{19} \,\text{eV}$
or more will be available in the future and that they come from all
directions in space \cite{BhattacharjeeSigl1998,Stanev2004,Armengaud2005}.
Throughout this article (except in the last equation of this Appendix),
we take the primary particle of the UHECR considered
to be a proton, but for a primary nucleus we can simply
replace the proton mass $M_\text{p}$ in \eqref{eq:generalbound}
by the relevant mass $M_\text{nucleus}$.
(The primary-particle type is expected to correlate
with, e.g., the atmospheric depth of the air-shower maximum;
cf. Ref.~\cite{Risse-etal2004}.)
With these assumptions, only three simple cases will be discussed in detail,
leaving a complete discussion to the moment when the radiation
rate \eqref{eq:other-dWdt} has been calculated exactly.
By the way, the \emph{implicit} assumption
behind the bounds of this Appendix is that no
extremely small numerical factors
appear in the final version of the radiation rate
\eqref{eq:other-dWdt} and that, therefore, the Cherenkov threshold
condition \eqref{eq:generalbound} is relevant;
cf. Ref.~\cite{ColemanGlashow1997}.

The first case considered has all space-space components of
$\widetilde{\kappa}_{\mu\nu}$ vanishing,
\begin{equation}\label{eq:case1}
\text{case\,1}:\;\;\widetilde{\kappa}_{ij}\equiv 0,
\;\;\text{for}\;\;i,j \in \{1,2,3\},
\end{equation}
with $\widetilde{\kappa}_{00} \equiv \widetilde{\kappa}_{jj}$
also vanishing.
Then, only the linear $\widehat{\vec q}$ term
on the left-hand-side of \eqref{eq:generalbound} remains.
Assuming to have observed $10\,\text{EeV}$ protons with
directions $\widehat{\vec q}$ $=$ $(\pm 1,0,0)$, $(0,\pm 1,0)$,
and $(0,0,\pm 1)$, the following bounds are obtained:
\begin{equation}\label{eq:case1-bounds}
\text{case\,1}:\;\; |\widetilde{\kappa}_{0j}|
\lesssim (1/2) \times 10^{-20},
\end{equation}
for $j = 1,2,3$.

The second case has spatial isotropy for the terms on the
diagonal of the $\widetilde{\kappa}_{\mu\nu}$ matrix
and vanishing off-diagonal terms:
\begin{equation}\label{eq:case2}
\text{case\,2}:\;\;
\widetilde{\kappa}_{11}\equiv \widetilde{\kappa}_{22} \equiv
\widetilde{\kappa}_{33}\equiv \widetilde{\kappa}_{00}/3\;\;
\text{and}\;\;
\widetilde{\kappa}_{\mu\nu}\equiv 0,
\;\;\text{for}\;\;\mu\ne\nu.
\end{equation}
The largest component of $\widetilde{\kappa}_{\mu\nu}$
is $\widetilde{\kappa}_{00}$.
Assuming to have observed at least one $300\,\text{EeV}$
proton \cite{Bird-etal1995,Risse-etal2004}, its particular
direction $\widehat{\vec q}$ being irrelevant, the following
bound is obtained for nonnegative $\widetilde{\kappa}_{00}$:
\begin{equation}\label{eq:case2-newbounds}
\text{case\,2}:\;\; 0\leq
\widetilde{\kappa}_{00} \lesssim (3/4)\times 10^{-23},
\end{equation}
having used a numerical value $E_\text{p}=300\,\text{EeV}$
for the right-hand-side of \eqref{eq:generalbound}.
There is no Cherenkov bound for negative $\widetilde{\kappa}_{00}$
and the previous electron-anomalous-magnetic-moment bound
(at the two $\sigma$ level) is:
\begin{equation}\label{eq:case2-oldbound}
\text{case\,2}:\;\;
0 < -\widetilde{\kappa}_{00}
\lesssim (3/2)\times 3 \times 10^{-8}
\approx   5 \times 10^{-8},
\end{equation}
which follows from Eq. (2.7) of
Ref.~\cite{Carone-etal2006}
with an absolute value on its left-hand-side.

The third case is based on a preferred background vector
$\boldsymbol{\xi}$, which, without loss of generality,
can be chosen as $\boldsymbol{\xi}\equiv (0,0,1)$.
The components of $\widetilde{\kappa}_{\mu\nu}$ are then given by
\begin{equation}\label{eq:case3-components}
\text{case\,3}:\;\;
\widetilde{\kappa}_{00}\equiv -\widetilde{\kappa}_{11} \equiv
-\widetilde{\kappa}_{22}\equiv \widetilde{\kappa}_{33}/3\;\;
\text{and}\;\;
\widetilde{\kappa}_{\mu\nu}\equiv 0,
\;\;\text{for}\;\;\mu\ne\nu.
\end{equation}
For this case, the largest component of $\widetilde{\kappa}_{\mu\nu}$
is $\widetilde{\kappa}_{33}$. Denoting the angle between the particle
momentum $\vec q$ and background vector $\boldsymbol{\xi}$ by
$\theta_3$,
the Cherenkov threshold condition \eqref{eq:generalbound} reads
\begin{equation}\label{eq:case3-angle}
\text{case\,3}:\;\; 0\leq
(4/3) \,\widetilde{\kappa}_{33}\, \cos^2\theta_3 \lesssim  10^{-20},
\end{equation}
for the numerical values of $E_\text{p}$ and $M_\text{p}$ used
on the right-hand-side of \eqref{eq:generalbound}.
Assuming the observation of a $10\,\text{EeV}$
cosmic proton moving parallel or antiparallel to the background
vector $\boldsymbol{\xi}$ in the 3--direction
(this proton having $\cos^2\theta_3=1$),
the following bound is obtained for nonnegative
$\widetilde{\kappa}_{33}$:
\begin{equation}\label{eq:case3-newbounds}
\text{case\,3}:\;\; 0\leq
 \widetilde{\kappa}_{33} \lesssim (3/4) \times 10^{-20}.
\end{equation}
Similar to case 2, there is no Cherenkov bound for
negative $\widetilde{\kappa}_{33}$ and the
previous microwave-oscillator bound (at the two $\sigma$ level) is:
\begin{equation}\label{eq:case3-oldbound}
\text{case\,3}:\;\;
0 < -\widetilde{\kappa}_{33}
\lesssim (9/4)\times (501/100) \times 10^{-14}
\approx   1.1 \times 10^{-13},
\end{equation}
which follows from Table II of
Ref.~[38(a)]  
with the identification
$\widetilde{\kappa}_{e-}=(4/9)\,\mathrm{diag}(1, 1, -2)\,\widetilde{\kappa}_{33}$
for the case considered with $\boldsymbol{\xi}\equiv (0,0,1)$ in
arbitrary coordinates.

In view of the above results, three general remarks are in order.
First, the overall sign of the $\kappa$--tensor in
\eqref{eq:modM-action} is physically relevant,
as the Cherenkov threshold condition \eqref{eq:generalbound} makes clear.
Second, cases 2 and 3 arise from a common \emph{Ansatz}:
$\widetilde{\kappa}_{\mu\nu} \propto \xi_\mu \xi_\nu
- \eta_{\mu\nu} \,\xi_\rho \xi^\rho  \, / 4$,
with $\xi^\mu=(1,0,0,0)$ for case 2 and
$\xi^\mu=(0,\boldsymbol{\xi})=(0,0,0,1)$ for case 3.
Third,
case 2 with negative $\widetilde{\kappa}_{00}$
and case 3 with negative $\widetilde{\kappa}_{33}$
have phase and group velocities of light
larger than the velocity $c$ encoded in
the causal structure of Minkowski spacetime (recall $x^0=c\,t$)
and it is not clear if these theories are physically consistent;
cf. Refs.~\cite{AdamKlinkhamer2001,ColladayKostelecky1998}.
Admittedly, case 1 is not perfect either, with
velocities of  light larger than $c$ in certain directions.
But case 2 with nonnegative $\widetilde{\kappa}_{00}$
(or case 3 with nonnegative $\widetilde{\kappa}_{33}$)
does appear to be physical and precisely the isotropic
case 2 with nonnegative $\widetilde{\kappa}_{00}$ has the
strongest bound, namely, Eq.~\eqref{eq:case2-newbounds}.

The three special cases discussed here give an idea of what a bound of
the type \eqref{eq:generalbound} may or may not imply for the coupling
constants $\widetilde{\kappa}_{\mu\nu}$.
Now consider generic $\widetilde{\kappa}_{\mu\nu}$
components with a \emph{positive}
left-hand-side in \eqref{eq:generalbound} of
order $\widetilde{\kappa}$, which is, most likely,
one of the conditions defining
the ``physical domain'' of modified-Maxwell theory \eqref{eq:modM-action}
coupled to Lorentz-invariant charged particles.
Then, the observation of a particular $300\,\text{EeV}$
proton \cite{Bird-etal1995,Risse-etal2004}
gives the tentative bound discussed in Sec.~\ref{sec:modM-Cherenkov}:
\begin{equation}\label{eq:genericUHECRbound}
0 \leq \widetilde{\kappa}\,
\big|^{\text{generic}\;\widetilde{\kappa}_{\mu\nu}}_\text{physical domain}
\lesssim 10^{-23}\,
\left(\frac{3 \times 10^{11} \, \text{GeV}}{E_\text{p}}\right)^2
\,\left(\frac{M_\text{p}}{\text{GeV}/c^2}\right)^2 \,,
\end{equation}
with the  above-mentioned \emph{caveat} on the
nature of the primary [for an iron (Fe) nucleus, the bound is increased by
a factor $(56)^{2}\approx 3 \times 10^{3}$ to an approximate value
of $3 \times 10^{-20}\,$].
Bound \eqref{eq:genericUHECRbound} is called tentative because
a \emph{precise} definition of ``generic $\widetilde{\kappa}_{\mu\nu}$''
has not been given.

However, with the reconstructed shower path available \cite{Bird-etal1995},
the proton direction of this particular $300\,\text{EeV}$ event ($n=1$) is
known within certain errors, $\widehat{\vec q}$ $=$ $\widehat{\vec q}^{(1)}$,
and can simply be inserted on the left-hand-side of \eqref{eq:generalbound},
with the right-hand-side taking the numerical value $10^{-23}$.
If more $300\,\text{EeV}$ protons become
available in the future ($n=1,\cdots,N$), expression \eqref{eq:generalbound}
generates $N$ bounds with
$\widehat{\vec q}$ $=$ $\widehat{\vec q}^{(n)}$ on the left-hand-side
and $10^{-23}$ on the right-hand-side, which can then be
analyzed further (giving proper confidence limits, for example).

The following \emph{Ansatz} for
the coupling constants $\widetilde{\kappa}^{\mu\nu}$
may turn out to be useful in the analysis:
\newcommand{\third}{\textstyle{\frac{1}{3}}}
\begin{equation}\label{eq:widetilde-kappa-mu-nu-Ansatz}
\big(\widetilde{\kappa}^{\mu\nu}\big) \equiv
\text{diag}\big(1,\third,\third,\third\big)\,
\overline{\kappa}^{00}
+\big(\delta\widetilde{\kappa}^{\mu\nu}\big),\;\;
\delta\widetilde{\kappa}^{00}=0,
\end{equation}
with one independent variable $\overline{\kappa}^{00}$
for the ``spatially isotropic part'' of $\widetilde{\kappa}^{\mu\nu}$
and eight independent variables in $\delta\widetilde{\kappa}^{\mu\nu}$
which need not be small. For a large number $N$
of $300\,\text{EeV}$ protons distributed
isotropically and restricting to the physical domain of the coupling
constants, the \emph{sum} of the $N$ bounds mentioned in the previous
paragraph will give approximately the same bound on the isotropy variable
$\overline{\kappa}^{00}$
as in \eqref{eq:case2-newbounds} for case 2 above,
but now without assumptions on the other coupling constants
$\delta\widetilde{\kappa}^{\mu\nu}$.

Returning to the present situation, there is already the following
bound on nonnegative $\overline{\kappa}^{00}$ from the current number
of more or less isotropic $10\,\text{EeV}$ cosmic-ray events (notably
from AGASA and HiRes \cite{BhattacharjeeSigl1998,Stanev2004}
in the northern hemisphere and from preliminary data of Auger
[24(b)] 
in the southern hemisphere):
\begin{equation}\label{eq:overline-kappa00-bound}
0\leq \overline{\kappa}^{00}\,\big|_\text{physical domain}
\lesssim
2 \times 10^{-17}\,
\left( \frac{M_\text{primary}}{M_\text{Fe}}\right)^{2},
\end{equation}
where, most likely, light primaries can be selected by appropriate
cuts on the shower-maximum depth and other characteristics.
Here, a relatively narrow energy band around $10\,\text{EeV}$
has been considered, but, more generally,
bound \eqref{eq:overline-kappa00-bound} scales
as the inverse square of the average cosmic-ray energy.
Incidentally, the reference frame in which
bound \eqref{eq:overline-kappa00-bound}
holds is the one in which the cosmic-ray energies are measured.
This new astrophysics bound on the spatially isotropic part of the
coupling constants $\widetilde{\kappa}^{\mu\nu}$ at the $10^{-17}$ level
(or at the $10^{-20}$ level for proton primaries)
improves significantly upon the direct laboratory bound at the $10^{-7}$
level [38(b)]  
or the indirect laboratory bound at the $10^{-8}$ level \cite{Carone-etal2006}.

\end{appendix}


\newpage
\begin{thebibliography}{99}

\bibitem{Cherenkov1934}
P.A. Cherenkov,
``The visible glow of pure liquids under the action of $\gamma$-rays,''
Dokl. Akad. Nauk Ser. Fiz. {\bf 2}, 451 (1934).

\bibitem{Vavilov1934}
S.I. Vavilov,
``On the possible causes of the blue $\gamma$-glow in liquids,''
Dokl. Akad. Nauk Ser. Fiz. {\bf 2}, 457 (1934).

\bibitem{Cherenkov1937}
P.A. Cherenkov,
``Visible radiation produced by electrons moving in a medium with
velocities exceeding that of light,''
Phys.\ Rev.\  {\bf 52}, 378 (1937).


\bibitem{FrankTamm1934}
I.M. Frank and I.E. Tamm,
``Coherent visible radiation of fast electrons passing through matter,''
Dokl. Akad. Nauk Ser. Fiz. {\bf 14}, 109 (1937).

\bibitem{Ginzburg1940}
V.L. Ginzburg,
``Quantum theory of light radiation from an electron moving uniformly in a medium,''
Zh. Eksp. Teor. Fiz. {\bf 10}, 589 (1940).

\bibitem{Cox1944}
R.T. Cox,
``Momentum and energy of photon and electron in the \v{C}erenkov radiation,''
Phys. Rev. {\bf 66}, 106 (1944).

\bibitem{Frank1984}
I.M. Frank,
``A conceptual history of the Vavilov--Cherenkov radiation,''
Sov. Phys. Usp. {\bf 27}, 385 (1984)
[Uspekhi Fiz. Nauk. {\bf 143}, 111 (1984)].


\bibitem{Jelley1958}
J.V. Jelley,
\emph{\v{C}erenkov Radiation and Its Applications}
(Pergamon Press, London, 1958).

\bibitem{Zrelov1970}
V.P. Zrelov,
\emph{Cherenkov Radiation in High-Energy Physics, Volume I}
(Israel Program Scientific Translations, Jerusalem, 1970).

\bibitem{Afanasiev2004}
G.N. Afanasiev,
\emph{Vavilov--Cherenkov and Synchrotron Radiation: Foundations and Applications}
(Kluwer Academic, Dordrecht, 2004).

\bibitem{Einstein1905}
A. Einstein,
``On the electrodynamics of moving bodies,''
Ann. Phys.\ (Leipzig)  {\bf 17}, 891 (1905); reprinted in
Ann. Phys.\ (Leipzig)  {\bf 14}, Suppl., 194 (2005).

\bibitem{Sommerfeld1904}
A. Sommerfeld,
``On the theory of electrons I, II, III,''
G\"{o}tt. Nachr. {\bf 2}, 99 (1904); {\bf 2}, 363 (1904);
{\bf 3}, 201 (1905).
In these prerelativity papers, Sommerfeld calculated, in particular,
the radiation loss of a hypothetical electron moving with constant
velocity $v>c$ in a classical vacuum where
electromagnetic waves propagate according to Maxwell theory
($c$ being the speed of light \emph{in vacuo}).
His Eq.~(65) of Part II, for example,
is to be compared to Eq.~(2.18) in Ref.~\cite{Jelley1958}, where the
last equation mentioned essentially equals
the Frank--Tamm formula \eqref{eq:radiated-classical}
integrated over the frequency range $[0,\omega_\text{c}]$,
for constant $n$ and cutoff $\omega_\text{c}\equiv c/(n\,d)$.
Apparently, Frank and Tamm \cite{FrankTamm1934} became aware of the
Sommerfeld papers only after they had completed their calculation.
For further details on the (pre)history of the Cherenkov effect,
see, e.g., Ref.~\cite{Frank1984}.

\bibitem{Beall1970}
E.F. Beall,
``Measuring the gravitational interaction of elementary particles,''
Phys.\ Rev.\ D {\bf 1}, 961 (1970), Sec. III A.  


\bibitem{ColemanGlashow1997}
S.R. Coleman and S.L. Glashow,
``Cosmic ray and neutrino tests of special relativity,''
Phys. Lett. B {\bf 405}, 249 (1997), hep-ph/9703240.

\bibitem{Carroll-etal1990}
  S.M. Carroll, G.B. Field, and R. Jackiw,
  ``Limits on a {L}orentz- and parity-violating
  modification of electrodynamics,''
  Phys. Rev. D {\bf 41}, 1231 (1990).

\bibitem{AdamKlinkhamer2001}
C. Adam and F.R. Klinkhamer,
``Causality and CPT violation from an Abelian Chern--Simons-like term,''
Nucl.\ Phys.\  B {\bf 607}, 247 (2001), hep-ph/0101087.

\bibitem{AdamKlinkhamer2003}
C. Adam and F.R. Klinkhamer,
``Photon decay in a CPT--violating extension of quantum electrodynamics,''
Nucl. Phys. B {\bf 657} (2003) 214, hep-th/0212028.

\bibitem{KaufholdKlinkhamer2006}
  C. Kaufhold and F.R. Klinkhamer, ``Vacuum Cherenkov radiation and photon
  triple-splitting in a Lorentz-noninvariant extension of quantum
  electrodynamics,''
  Nucl. Phys. B {\bf 734}, 1 (2006), hep-th/0508074.

\bibitem{LehnertPotting2004PRL93}
R. Lehnert and R. Potting,
``Vacuum \v{C}erenkov radiation,''
Phys.\ Rev.\ Lett.\  {\bf 93}, 110402 (2004), hep-ph/0406128.

\bibitem{LehnertPotting2004PRD70}
  R. Lehnert and R. Potting,
  ``\v{C}erenkov effect in Lorentz-violating vacua,''
  Phys. Rev. D {\bf 70}, 125010 (2004), hep-ph/0408285.


\bibitem{Wardle-etal1997}
J. Wardle, R. Perley, and M. Cohen,
``Observational evidence against birefringence over
  cosmological distances,''
Phys. Rev. Lett. {\bf 79}, 1801  (1997), astro-ph/9705142.

\bibitem{BhattacharjeeSigl1998}
P. Bhattacharjee and G. Sigl,
``Origin and propagation of extremely high energy cosmic rays,''
Phys.\ Rept.\  {\bf 327}, 109 (2000), astro-ph/9811011.


\bibitem{Stanev2004}
T. Stanev,
``Ultra high energy cosmic rays,''
in: \emph{Proceedings of 32nd SLAC Summer Institute on Particle Physics (SSI 2004):
Natures Greatest Puzzles}, eConf C040802, L020 (2004), astro-ph/0411113.


\bibitem{Armengaud2005}
(a) E. Armengaud  [Pierre Auger Collaboration],
``Search methods for UHECR anisotropies within the Pierre Auger Observatory,''
report FERMILAB-CONF-05-070-A-E-TD, April 2005;
(b) E. Armengaud,
``Propagation and distribution on the sky of ultra-high-energy cosmic
  rays for the Pierre Auger Observatory,''
  PhD thesis, Universit\'{e} Paris 7, May 2006.



\bibitem{Kostelecky2003}
V.A. Kosteleck\'{y},
``Gravity, Lorentz violation, and the standard model,''
Phys.\ Rev.\  D {\bf 69}, 105009 (2004), hep-th/0312310.

\bibitem{KantKlinkhamer2005}
E. Kant and F.R. Klinkhamer,
``Maxwell--Chern--Simons theory for curved spacetime backgrounds,''
Nucl.\ Phys.\  B {\bf 731}, 125 (2005), hep-th/0507162.


\bibitem{Carusotto-etal2001}
I. Carusotto, M. Artoni, G.C. La Rocca, and F. Bassani,
``Slow group velocity and Cherenkov radiation,''
Phys. Rev. Lett. {\bf 87}, 064801 (2001), quant-ph/0103059.

\bibitem{Luo-etal2001}
C. Luo, M. Ibanescu, S.G. Johnson, and J.D. Joannopoulos,
``\v{C}erenkov radiation in photonic crystals,''
Science {\bf 299}, 368 (2003).

\bibitem{PanofskyPhillips1962}
 W.K.H. Panofsky and M. Phillips,
\emph{Classical Electricity and Magnetism}, second edition
(Addison--Wesley, Reading, MA, 1962).

\bibitem{LandauLifshitz1984}
L.D. Landau, E.M. Lifshitz, and L.P. Pitaevskii,
\emph{Electrodynamics of Continuous Media}, second edition
(Butterworth--Heinemann, Oxford, 1984).

\bibitem{Jackson1999}
J.D. Jackson,
\emph{Classical Electrodynamics}, third edition
(Wiley, New York, 1999).



\bibitem{ChadhaNielsen1982}
S. Chadha and H.B. Nielsen,
``Lorentz invariance as a low-energy phenomenon,''
Nucl.\ Phys.\  B {\bf 217}, 125 (1983).


\bibitem{ColladayKostelecky1998}
D. Colladay and V.A. Kosteleck\'{y},
``Lorentz-violating extension of the standard model,''
Phys. Rev. D {\bf 58}, 116002 (1998), hep-ph/9809521.

\bibitem{Altschul2007PRL98}
B. Altschul,
``Vacuum \v{C}erenkov radiation in Lorentz-violating theories
without CPT violation,''
Phys. Rev. Lett. {\bf 98}, 041603 (2007), hep-th/0609030.

\bibitem{Altschul2007PRD75}
B. Altschul,
``\v{C}erenkov radiation in a Lorentz-violating and birefringent vacuum,''
Phys.\ Rev.\  D {\bf 75}, 105003 (2007), hep-th/0701270.


\bibitem{KosteleckyMewes2002}
V.A. Kosteleck\'{y} and M. Mewes,
``Signals for Lorentz violation in electrodynamics,''
Phys.\ Rev.\  D {\bf 66}, 056005 (2002), hep-ph/0205211.

\bibitem{Carone-etal2006}
C.D. Carone, M. Sher, and M. Vanderhaeghen,
``New bounds on isotropic Lorentz violation,''
Phys.\ Rev.\  D {\bf 74}, 077901 (2006), hep-ph/0609150.


\bibitem{Stanwix-etal2006}
(a)  P.L. Stanwix, M.E. Tobar, P. Wolf, C.R. Locke, and E.N. Ivanov,
``Improved test of Lorentz invariance in electrodynamics using rotating
cryogenic sapphire oscillators,''
Phys.\ Rev.\  D {\bf 74}, 081101 (2006), gr-qc/0609072;
(b)  M. Hohensee, A. Glenday, C.H. Li, M.E. Tobar, and P. Wolf,
``Erratum: New methods of testing Lorentz violation in electrodynamics,''
Phys.\ Rev.\  D {\bf75}, 049902 (2007), hep-ph/0701252.

\bibitem{Bird-etal1995}
D.J.  Bird et al.,
``Detection of a cosmic ray with measured energy well beyond the expected
spectral cutoff due to cosmic microwave radiation,''
Astrophys. J.  {\bf 441}, 144 (1995), astro-ph/9410067.

\bibitem{Risse-etal2004}
M. Risse,  P. Homola, D. Gora, J. Pekala, B. Wilczynska, and H. Wilczynski,
``Primary particle type of the most energetic Fly's Eye air shower,''
Astropart. Phys.  {\bf 21}, 479 (2004), astro-ph/0401629.


\bibitem{Klinkhamer2000}
F.R. Klinkhamer,
``A CPT anomaly,''
Nucl.\ Phys.\  B {\bf 578}, 277 (2000), hep-th/9912169.

\bibitem{KlinkhamerSchimmel2002}
F.R. Klinkhamer and J. Schimmel,
``CPT anomaly: A rigorous result in four dimensions,''
Nucl.\ Phys.\  B {\bf 639}, 241 (2002), hep-th/0205038.

\bibitem{Klinkhamer2005}
F.R. Klinkhamer,
``Nontrivial spacetime topology, CPT violation, and photons,''
in: \emph{CP Violation and the Flavour Puzzle:
Symposium in Honour of Gustavo C. Branco},
edited by D. Emmanuel-Costa et al.
(Poligrafia Inspektoratu, Krak\'{o}w, Poland, 2005),
pp. 157--191, hep-ph/0511030.


\bibitem{Collins-etal2004}
J. Collins, A. Perez, D. Sudarsky, L. Urrutia, and H. Vucetich,
``Lorentz invariance: An additional fine-tuning problem,''
Phys.\ Rev.\ Lett.\ {\bf 93}, 191301 (2004), gr-qc/0403053.

\bibitem{BernadotteKlinkhamer2007}
S. Bernadotte and F.R. Klinkhamer,
``Bounds on length scales of classical spacetime foam models,''
Phys.\ Rev.\  D {\bf 75}, 024028 (2007), hep-ph/0610216.

\bibitem{Wolfram99}
S. Wolfram,
\emph{Mathematica}
(Cambridge Univ. Press, Cambridge, UK, 1999).

\end{thebibliography}
\end{document}